\documentclass[reprint, superscriptaddress, amsmath, amssymb,
aps, showkeys, showpacs, twoside, final, secnumarabic,
nofootinbib]{revtex4-2}

\usepackage[paperwidth=205mm,paperheight=290mm,top=17mm,bottom=25mm,
inner=17mm,outer=17mm, twoside]{geometry}

\usepackage[english]{babel}
\usepackage{color}
\usepackage{graphicx}
\usepackage[unicode=true,colorlinks=true,linkcolor=magenta,
urlcolor=blue, citecolor = blue,breaklinks]{hyperref}

\usepackage{multirow}
\usepackage{breakurl}
\newcount\issue
\newcount\Vol
\newcount\numb
\usepackage{fancyhdr}
\def\Vol{\textbf{79}}
\def\numb{x}
\graphicspath{{figs/}}
\newcommand{\spb}{EUSO-SPB2}
\newcommand{\ta}{EUSO-TA}
\newcommand\tp{\ensuremath{\mathrm{TP}}}

\newcommand\fp{\ensuremath{\mathrm{FP}}}
\newcommand\fn{\ensuremath{\mathrm{FN}}}
\newcommand{\offline}{\mbox{$\overline{\rm Off}$\hspace{.05em}\raisebox{.3ex}{$\underline{\rm line}$}}}

\begin{document}

\title{Machine Learning in Fundamental Physics\\[20pt]
Reconstruction of energy and arrival directions of UHECRs registered
by~fluorescence telescopes with neural networks
}

\def\addressa{D.V.~Skobeltsyn Institute of Nuclear Physics,
M.V.~Lomonosov Moscow State University,\\ Moscow 119234, Russia}

\author{\firstname{Mikhail}~\surname{Zotov}}
\email[E-mail: ]{zotov@eas.sinp.msu.ru}
\affiliation{\addressa}
\collaboration{for the JEM-EUSO collaboration}

\received{xx.10.2024}
\revised{xx.xx.2024}
\accepted{xx.xx.2024}

\begin{abstract}

	Fluorescence telescopes are important instruments widely used in
	modern experiments for registering ultraviolet radiation from
	extensive air showers (EASs) generated by cosmic rays of ultra-high
	energies.  We present proof-of-concept convolutional neural networks
	aimed at reconstruction of energy and arrival directions of primary
	particles using model data for two telescopes developed by the
	international JEM-EUSO collaboration.  We also demonstrate how a
	simple convolutional encoder-decoder can be used for EAS track
	recognition.  The approach is generic and can be adopted for other
	fluorescence telescopes.

\end{abstract}

\pacs{96.50.S-, 95.85.Ry, 84.35.+i, 07.05.Mh}
\keywords{ultra-high energy cosmic rays, fluorescence telescope,
reconstruction, pattern recognition, convolutional neural
network\\[5pt]}


\maketitle
\thispagestyle{fancy}

\section{Introduction}\label{intro}

Registering fluorescence radiation of extensive air showers (EASs) with
dedicated telescopes in nocturnal atmosphere of Earth is an established
technique of studying ultra-high energy cosmic rays
($\gtrsim10^{18}~\text{eV}=1~\text{EeV}$)~\cite{dawson_etal}.
Fluorescence telescopes (FTs) play an important role in both leading
experiments in this field of astrophysics, the Pierre Auger
Observatory~\cite{auger-ft} and the Telescope Array~\cite{ta-ft}.  They
were also suggested to be used as orbital telescopes due to the
opportunity to increase the exposure of an experiment
dramatically~\cite{benson-linsley}.  The international JEM-EUSO
collaboration is implementing a step-by-step program to conduct a
full-scale orbital experiment aimed at solving the long-standing puzzle
of the nature and origin of the highest energy cosmic
rays~\cite{jemeuso-mission, jemeuso-instrument, jemeuso-science}.  Two
of the instruments already built are the small ground-based \ta{}
telescope~\cite{eusota2015, eusota2018, eusota-2024}, which operates at
the site of the Telescope Array experiment, and the \spb{} stratospheric
experiment, which included both fluorescence and Cherenkov
telescopes~\cite{spb2-2021, spb2-2023}.

Reconstruction of parameters of initial UHECRs based on the signal
registered by a monocular FT is not a trivial
task~\cite{jemeuso-reco, jemeuso-energy, jemeuso-angular}. Here we
suggest two simple artificial neural networks that can be used to
recognize tracks registered by a single FT and to reconstruct energy and
arrival directions (ADs) of ultra-high energy primaries.  We employ data
simulated for \ta{} and the fluorescence telescope of \spb{} to train
and test these networks.  In what follows, we present preliminary
results of this technique and outline possible ways of its improvement.
We believe that the suggested approach is generic and can be adopted for
other fluorescence telescopes operating nowadays as well as the future
ones. It can also be modified to use benefits provided by stereoscopic
observations.

\section{Fluorescence telescopes EUSO-SPB2 and EUSO-TA}

EUSO-SPB2 (the Extreme Universe Space Observatory on a Super Pressure
Balloon~2) was a stratospheric pathfinder mission for a future orbital
experiment like POEMMA~\cite{poemma}. The scientific equipment included
two telescopes with an identical modified Schmidt
design of optics with a 1~m diameter entrance pupil occupied by an
aspheric corrector plate, a curved focal surface (FS), and a spherical
primary mirror. 
The Cherenkov telescope was designed to observe Cherenkov emission
of cosmic ray EASs with energies above 1~PeV. It could be tilted up and
down between~$3.5^\circ$ and $-13^\circ$ w.r.t.\ the horizon pointing
towards the limb~\cite{spb2-ct-2021, spb2-ct-2023}.
The fluorescence telescope was intended to register the fluorescence
emission of extensive air showers produced by UHECRs with the energy above
2~EeV. The telescope was pointed in the nadir direction and had a total
field of view (FoV) of $12^\circ\times36^\circ$.  The FT camera
consisted of three photo detector modules (PDMs), each built of
$48\times48$ pixels in a rectangular grid, thus giving 6912 pixels in
total. Every PDM was covered with a BG3 filter to limit its sensitivity
to the near-UV range (290--430~nm). The time resolution was equal to
1~$\mu$s~\cite{spb2-ft-2021a, spb2-ft-2023}.

\spb{} was launched on May 13th, 2023, from Wanaka, New Zealand, as a
mission of opportunity on a NASA super pressure balloon. It was expected
that the mission will continue up to 100 days and will result in the
first observation of UHECRs in the nadir direction. Unfortunately,
the mission terminated in approximately 37~hours after the start due to
a leak in the balloon. No EAS were found in the data transmitted to the
data center, which agrees with the expected trigger rate. Thus, we only
work with simulated data in the paper.

\ta{} is a ground-based fluorescence telescope built to serve as a
test-bed for testing the design, electronics, software and other aspects
of the future orbital missions~\cite{eusota2015, eusota2018}. It
operates at the site of the Telescope Array experiment in Utah, USA,
near its Black Rock Mesa FTs. \ta{} is a refractor-type telescope
consisting of two Fresnel lenses and concave focal surface. The lenses
have a diameter of 1~m and are manufactured of 8-mm thick PMMA. The FS
has the size of about $17~\text{cm}\times17~\text{cm}$ and consists of
$48\times48$ pixels, similar to one of the PDMs of \spb. The field of
view of one pixel equals $0.2^\circ\times0.2^\circ$ providing
approximately $10.6^\circ\times10.6^\circ$ in total. In what follows, we
are using time resolution of the detector equal to 2.5~$\mu$s but it is
expected that the electronics of the instrument will be upgraded to have
a 1~$\mu$s resolution, as that of the \spb{} FT.

\ta{} can operate at different elevation angles. In what follows, the
elevation angle of $10^\circ$ was used. Studies for different other
elevation angles are to be performed in the future.

\section{Simulated data sets}

We used CONEX~\cite{conex} to simulate EAS from UHECRs in the energy
range from a few EeV up to 100~EeV. The response of both instruments was
simulated with the \offline{} framework~\cite{offline}. For \ta, we
demanded that EAS cores were within the projection of the telescope FoV
on the ground.  The distance between the telescope and shower cores
varied in the range 2--40~km depending on the energy of primary protons.
For both instruments, we used events that flagged software triggers of
the respective telescope.  Out of these events, we excluded those that
contained just a few hit pixels in one of the corners of the field of
view of a telescope and less than two time frames with hit
pixels.\footnote{The right panel of Fig.~\ref{ta-angsep-out} provides an
example of a signal that was considered for reconstruction despite of
the small number of hit pixels.} No other quality cuts on the signals
were implied. In particular, we did not select events with the shower
maximum being in the FoV of a telescope, which can considerably improve
the quality of reconstruction.

We were primarily interested in the development of a neural network for
reconstructing energy of primary particles. By this reason, our main
data set used for training and testing the neural networks was prepared
with a quasi-uniform distribution of events wrt.\ energy: showers were
simulated with the step of 1~EeV in the range 10--30~EeV, and with the
step of 2~EeV for higher energies.\footnote{However, our tests
have demonstrated that a dataset simulated with a uniform distribution
of events wrt.\ $\lg E$ acts almost equally well in terms of the quality
of the energy reconstruction.}
Simulated events had a uniform distribution of azimuth angles. Zenith
angles were distributed $\sim\cos\theta$ as is the default in CONEX.
This might be sub-optimal for data sets aimed at training an artificial
neural network
aimed at reconstruction of UHECR arrival directions.
In what follows, all simulations were performed for proton primaries
with the QGSJETII-04 model of hadronic interactions~\cite{qgsjet}.
Simulations for heavier primaries and other models are possible in the
future.

The background illumination at the rate of 1 photon/pixel/time step was
simulated for both telescopes. This is the expected level of background
illumination during normal operation of the instruments in moonless
nights. However, it is ignored in the next section because we were
firstly interested in developing a proof-of-concept method rather than a
production-level one.

\section{Reconstruction of energy and arrival directions}

To perform reconstruction, we used a simple convolutional neural network
(CNN) similar to the one presented in~\cite{ml4spb2-icrc2023}. It
consists of six convolutional layers (CLs) with a maxpooling layer coming
after each even convolutional layer. Each of CLs
employs 36 filters with the kernel size of $4\times4$; the L2 kernel
regularizer is used.  Three fully connected layers with 512, 256, and
128 nodes come next.  Finally, there comes a layer with the number of
nodes equal to the number of reconstructed parameters. Adam is used as
an optimizer, ReLU is the activation function. The learning rate was
chosen automatically according to the behavior of the model loss, with
the initial rate typically equal to~$10^{-4}$.

The loss function was chosen according to the type of parameters to be
reconstructed. If energy was the only reconstructed parameter, we
employed the mean absolute percentage error (MAPE). If the list of
parameters included azimuth and/or zenith angles, the loss function was
either mean squared error (MSE) or mean absolute error (MAE).
In case we were only interested in
reconstruction of arrival directions, we also used angular separation as
the model metric. The coefficient of determination~$R^2$ was used as an
additional metric for evaluating the quality of models on test samples.

\subsection{EUSO-SPB2}

\begin{figure}[!ht]
	\includegraphics[width=.245\textwidth]{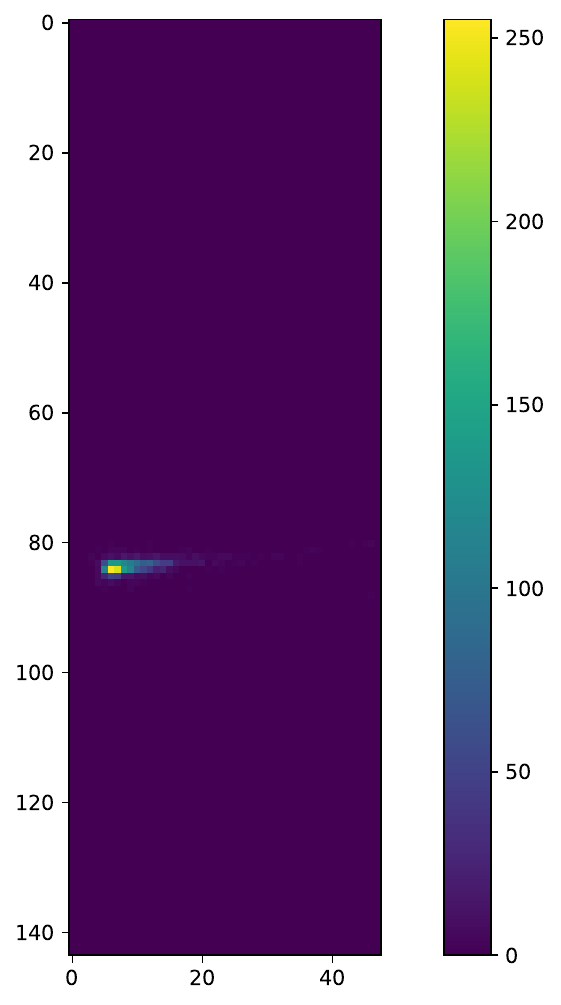}%
	\includegraphics[width=.24\textwidth]{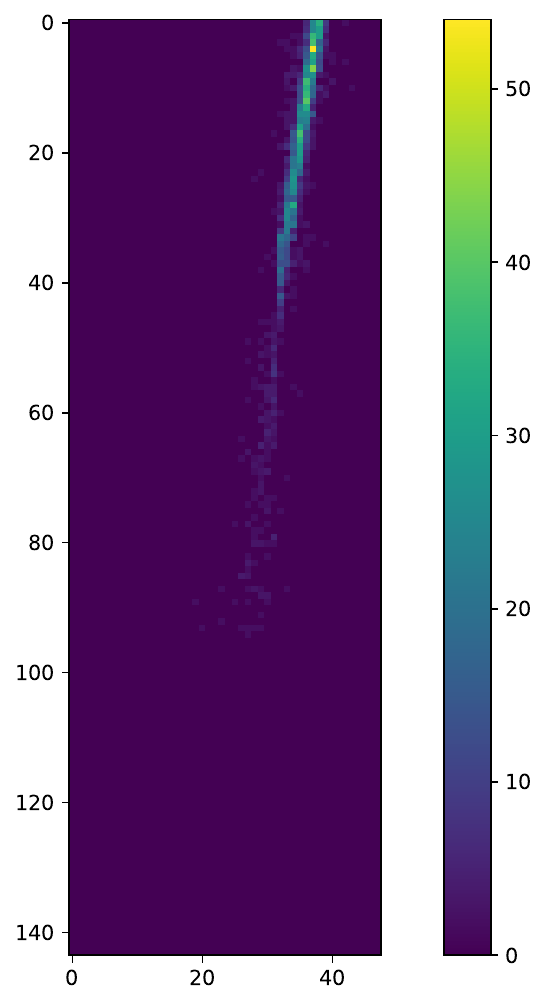}

	\caption{Examples of integral tracks of simulated EASs produced by
	UHECRs as seen by \spb.
	A track shown in the left panel was produced by a shower initiated by
	a 96~EeV proton arriving at the zenith angle $\theta=19^\circ$.
	The right panel shows a track from a 26~EeV proton arriving at
	$\theta=63^\circ$.
	Here and below colors indicate photon counts per
	pixel. Numbers along the axes represent pixel
	numbers.}

	\label{spb-examples}
\end{figure}

For \spb, we used a training data set consisting of nearly 50 thousand
samples with 20\% of them acting as a validation set.  Each sample
presented just an integral track of an event with pixels having photon
counts $<2$ zeroed because they won't be recognized anyway with the
expected level of the background illumination.  Fig.~\ref{spb-examples}
presents two examples of such tracks, with more examples below.

\begin{figure}[!ht]
	\includegraphics[width=.48\textwidth]{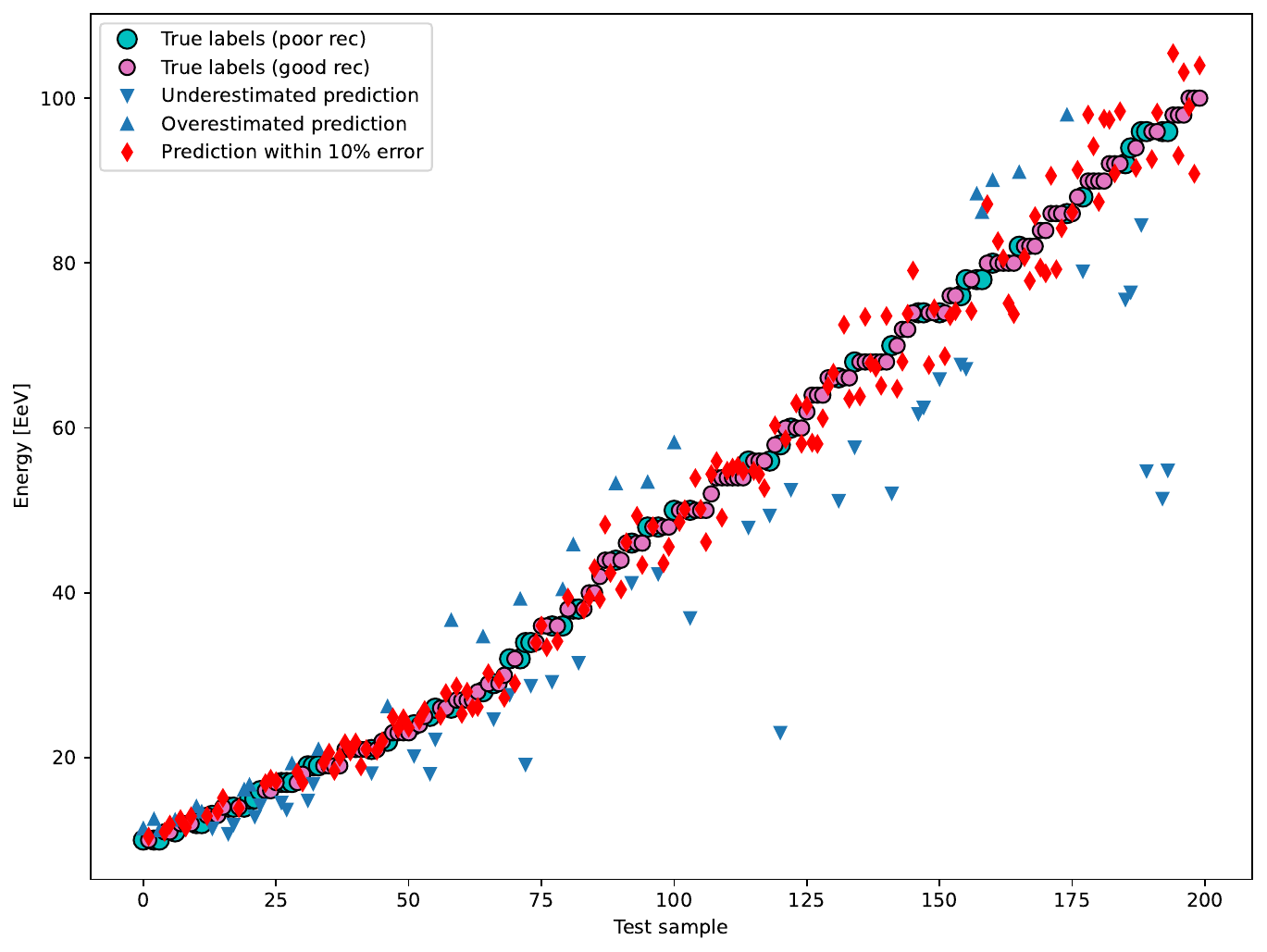}
	\caption{An example of energy reconstruction for \spb. Circles denote
	ground truth labels. Red diamonds indicate predictions that deviate
	form true labels by less than~10\%. Triangles show predicted energies
	with errors $\ge10\%$.}
	\label{spb-energy}
\end{figure}

\begin{figure}[!ht]
	\includegraphics[width=.48\textwidth]{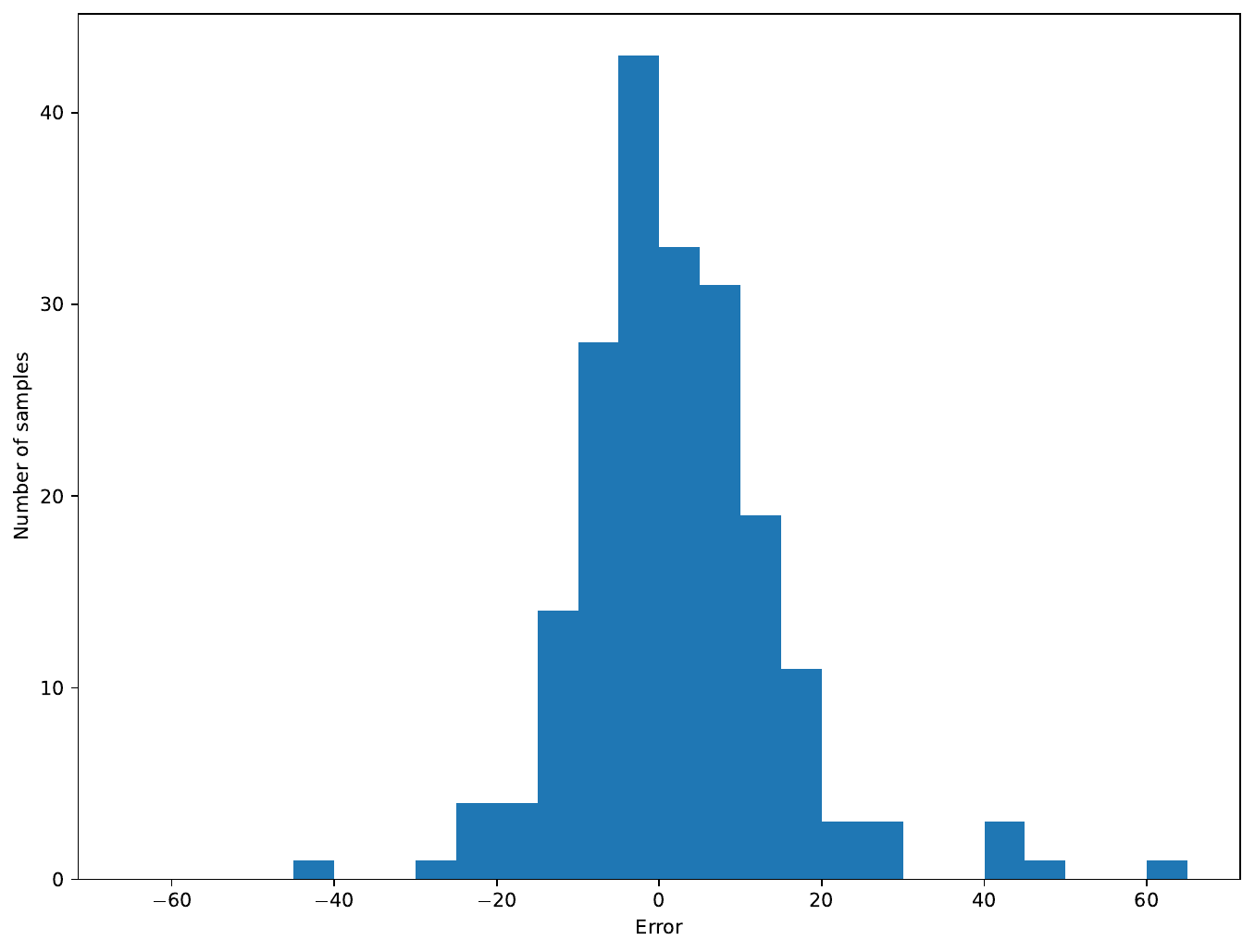}
	\caption{Histogram of errors expressed in percent for reconstruction
	of energy for \spb. The mean equals~2, the standard deviation equals
	12.6. Here and below error means the difference between
	true and predicted labels.}
	\label{spb-energy-hist}
\end{figure}

All data were scaled before feeding to the CNN in such a way that the
brightest pixel of the data set was equal to~1.  Figure~\ref{spb-energy}
presents results of energy reconstruction for a test sample consisting
of 200 events with energies from 10~EeV up to 100~EeV.
Figure~\ref{spb-energy-hist} demonstrates the distribution of errors
expressed in percent.  One can see that generally predictions follow the
ground truth labels with just a few outliers.  The mean absolute
percentage error equals 9.1\% with the maximum error equal to 60.2\%.
The mean error is~2\% with the standard deviation equal to 12.6. The
coefficient of determination $R^2=0.923$.

\begin{figure}[!ht]
	\includegraphics[width=.24\textwidth]{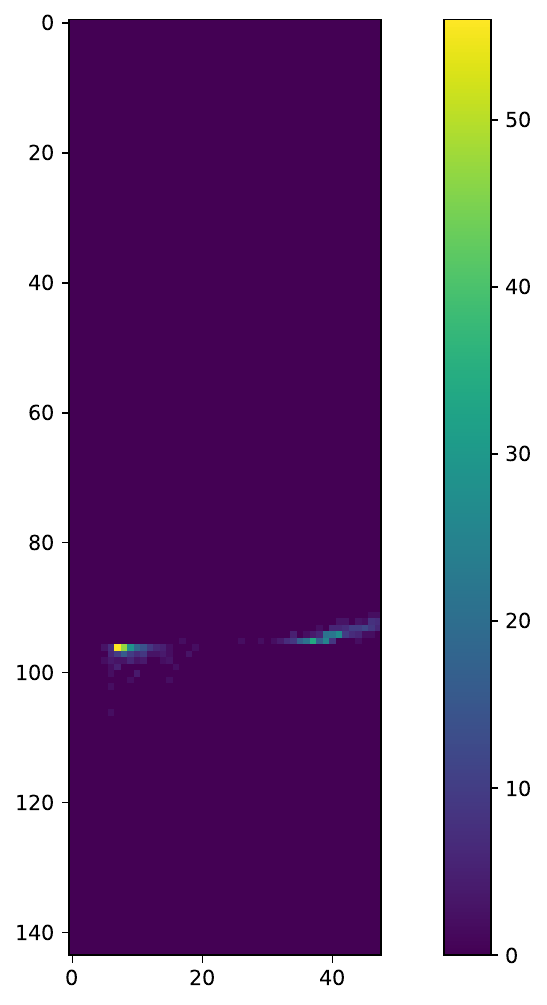}%
	\includegraphics[width=.24\textwidth]{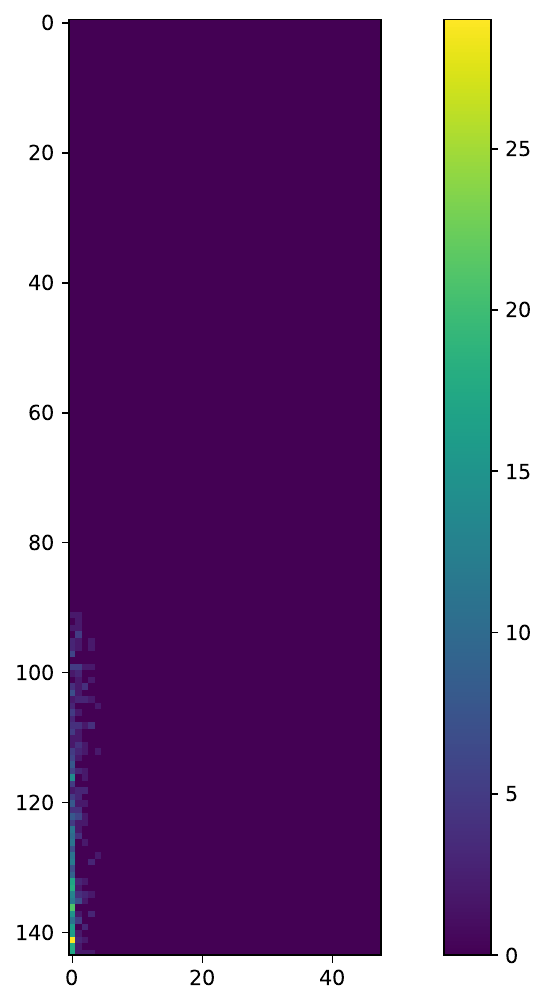}
	\caption{Integral tracks of two \spb{} events with strongly
	underestimated energies. See the text for details.}
	\label{spb-energy-out}
\end{figure}

One can see that outliers in Fig.~\ref{spb-energy} mostly have energies
strongly underestimated. Figure~\ref{spb-energy-out} shows a couple of
such outliers.
The left panel shows an event with true and predicted energies equal
to 58~EeV and 23~EeV respectively, thus resulting in a 60.2\% error.
One can see that the middle part of the track
is not registered at all. 
This is due to the missing part being located at the gap between two
PDMs, see Fig.~{3} in~\cite{spb2-ft-2021a}.
The track of the event shown in the
right panel goes exactly along the left edge of the focal surface so
that only a part of the illumination is registered in this case, too.
This results in a 44\% error with the true and predicted energies equal
to 34~EeV and 19.2~EeV respectively.

Energy of the event with a short track shown in the left panel of
Fig.~\ref{spb-examples} was also underestimated by~43\% while the error
for the long-track event shown in the right panel of the same figure was
only around~4\%.

It is worth mentioning that energy is reconstructed by the CNN without
any prior knowledge about arrival directions of primary particles. This is
in contrast with conventional algorithms that require the zenith angle
to be reconstructed first~\cite{jemeuso-angular}.

\begin{figure}[!ht]
	\includegraphics[width=.48\textwidth]{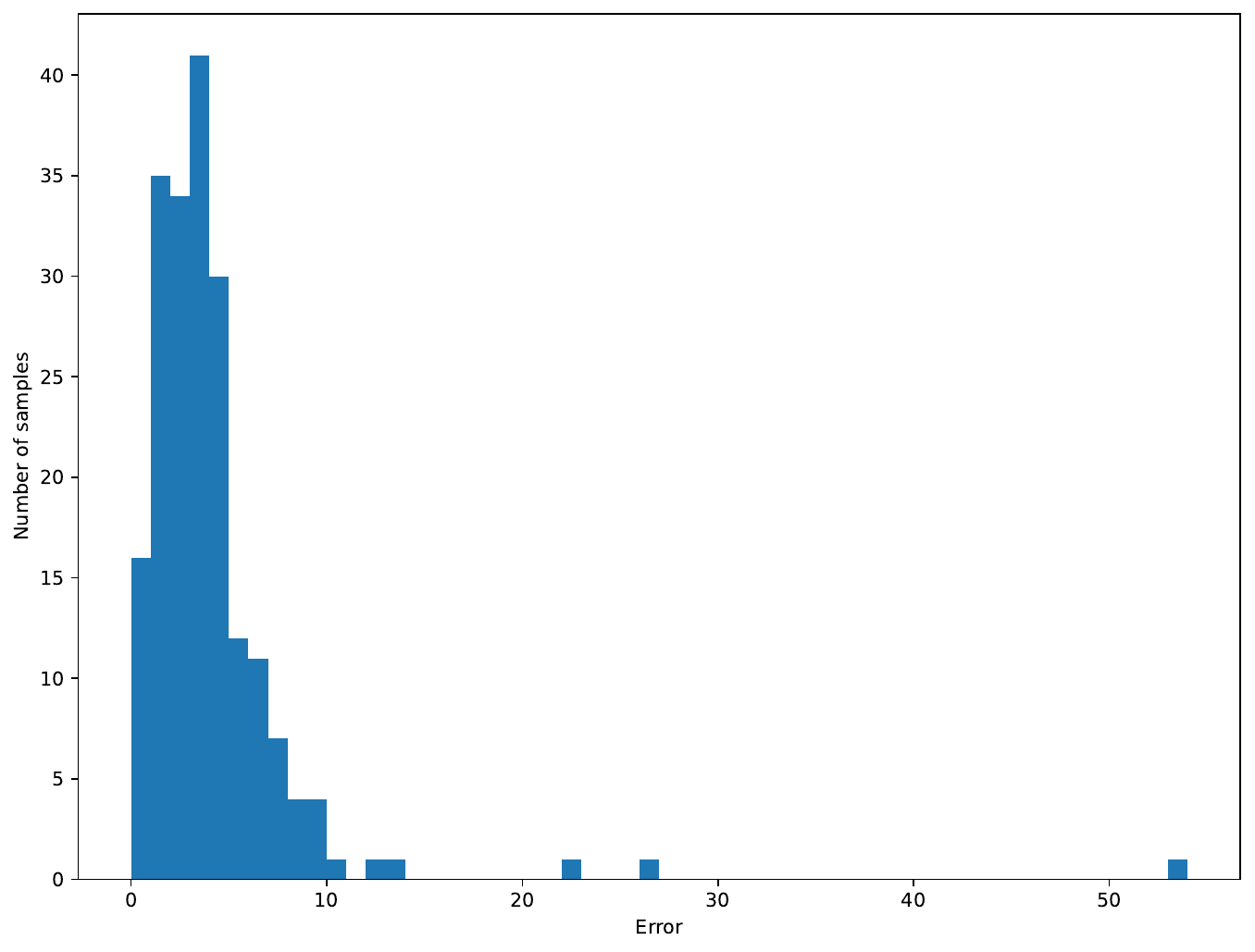}

	\caption{Distribution of errors in reconstruction of arrival
	directions for a test sample for \spb. Errors are expressed in degrees.
	See the text for details.}

	\label{spb-angsep}
\end{figure}

Now let us consider reconstruction of arrival directions of UHECRs.
Figure~\ref{spb-angsep} presents the distribution of errors in the
angular separation of true and predicted arrival directions for a test sample
consisting of 200 events. The mean angular separation equals~$4.1^\circ$
with the median equal to~$3.3^\circ$.
The coefficient of determination equals 0.948.\footnote{In some tests,
median of the angular separation was below~$1.5^\circ$ but we are
presenting a more typical result.}

\begin{figure}[!ht]
	\includegraphics[width=.48\textwidth]{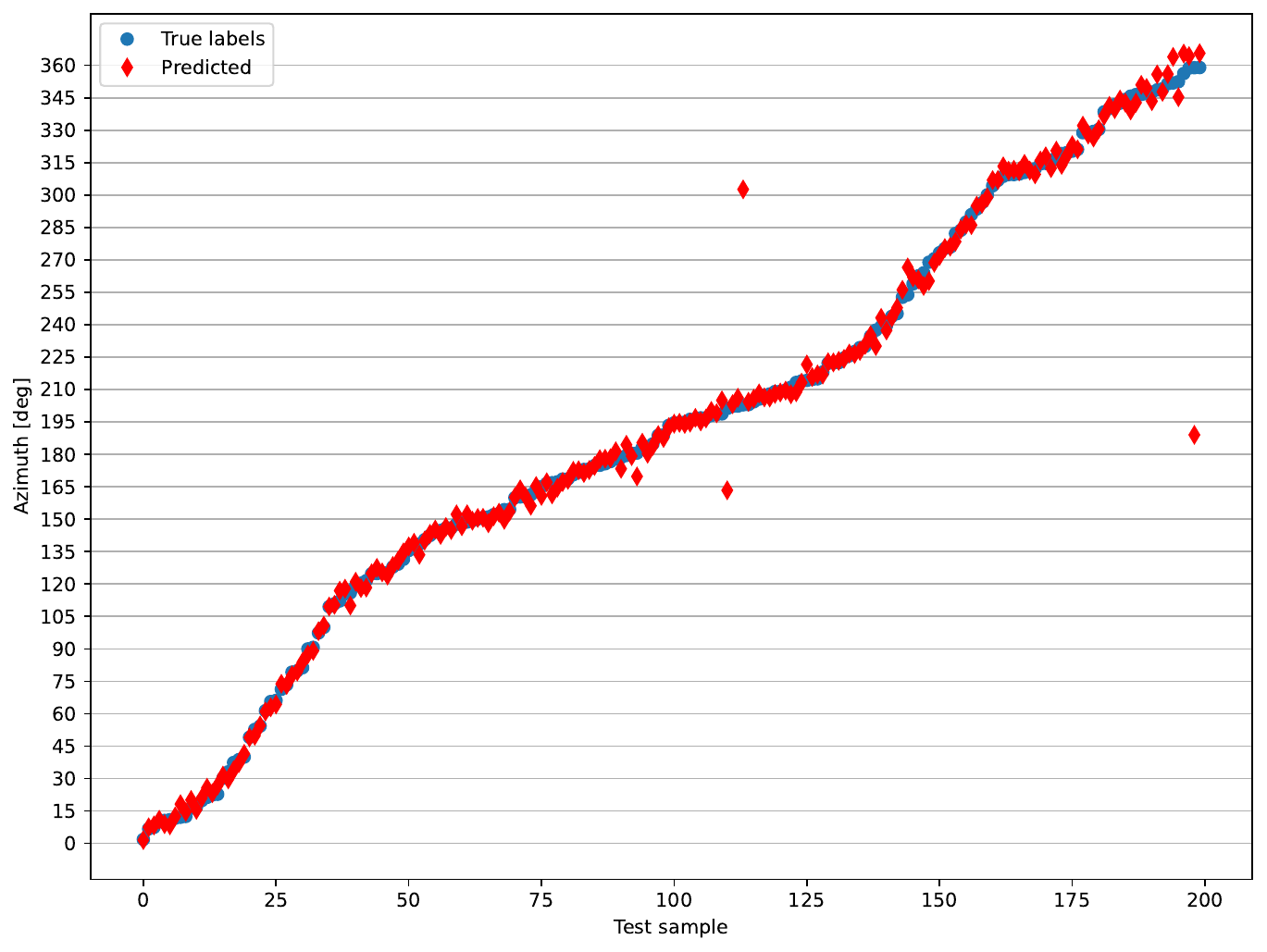}

	\caption{Reconstruction of azimuth angles for a test sample for \spb.
	Circles and diamonds indicate true and predicted values
	respectively.}

	\label{spb-azimuth}
\end{figure}

Our analysis reveals that poorly reconstructed ADs are due to large errors
in reconstruction of azimuth angles. This might come as a surprise
because azimuth angles seem to be easily estimated with a telescope
looking in nadir.  Figure~\ref{spb-azimuth} shows how azimuth angles
were reconstructed for the same test sample.

\begin{figure}[!ht]
	\includegraphics[width=.235\textwidth]{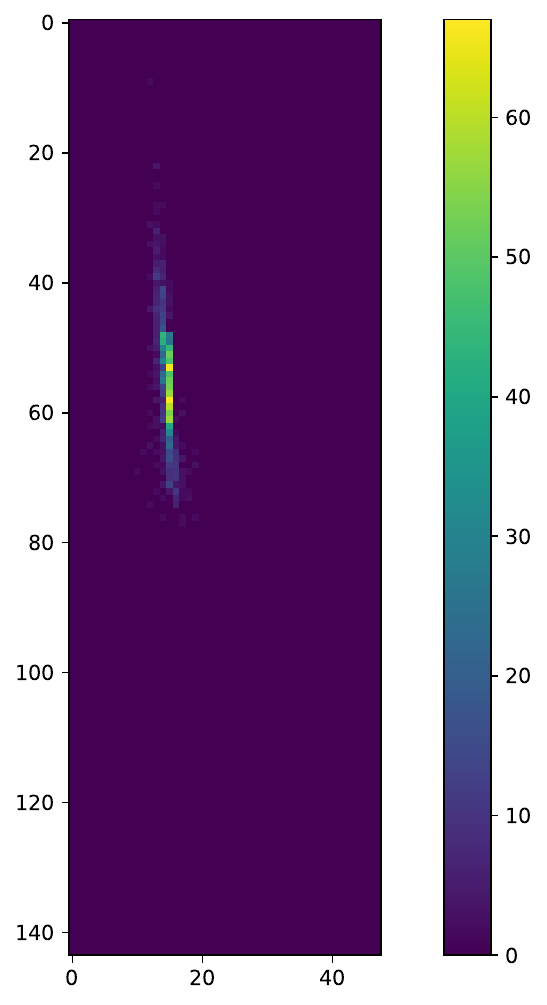}%
	\includegraphics[width=.24\textwidth]{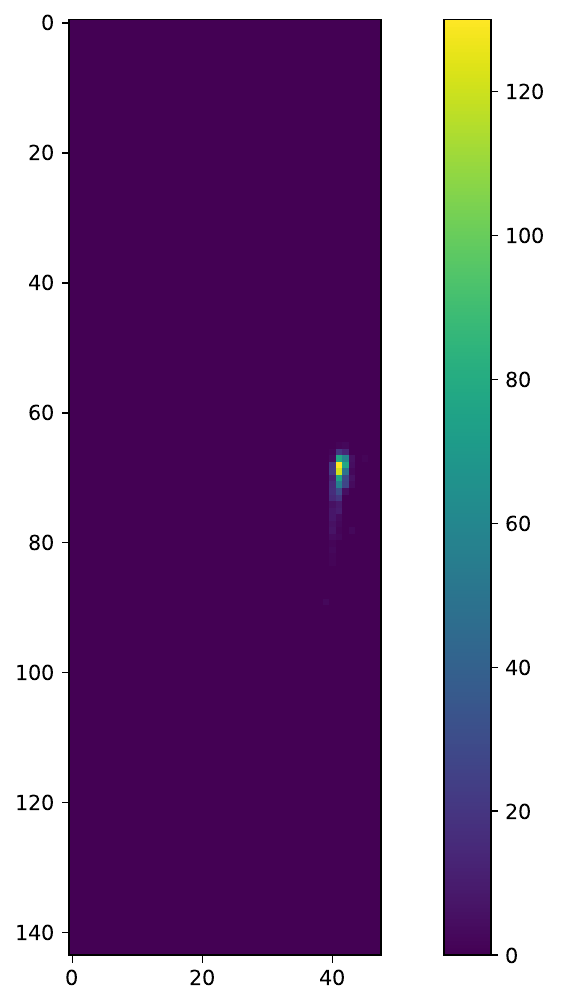}

	\caption{Integral tracks of two \spb{} events with the largest errors
	in reconstruction of arrival directions. See the text for details.}

	\label{spb-angsep-out}
\end{figure}

Let us consider two events with the largest errors in reconstructed
arrival directions. Figure~\ref{spb-angsep-out} presents tracks of these
outliers. For the event shown in the left panel, the CNN predicted an
almost opposite azimuth angle ($189.0^\circ$ instead of $358.9^\circ$)
thus resulting in the angular
separation between the true and predicted arrival directions equal
to~$53.9^\circ$.
The track shown in the right panel has a small footprint on the focal
surface seemingly not sufficient for an accurate reconstruction of the
azimuth angle. The angular separation between the true and predicted
arrival directions equals $26.8^\circ$ in this case.

\begin{figure}[!ht]
	\includegraphics[width=.48\textwidth]{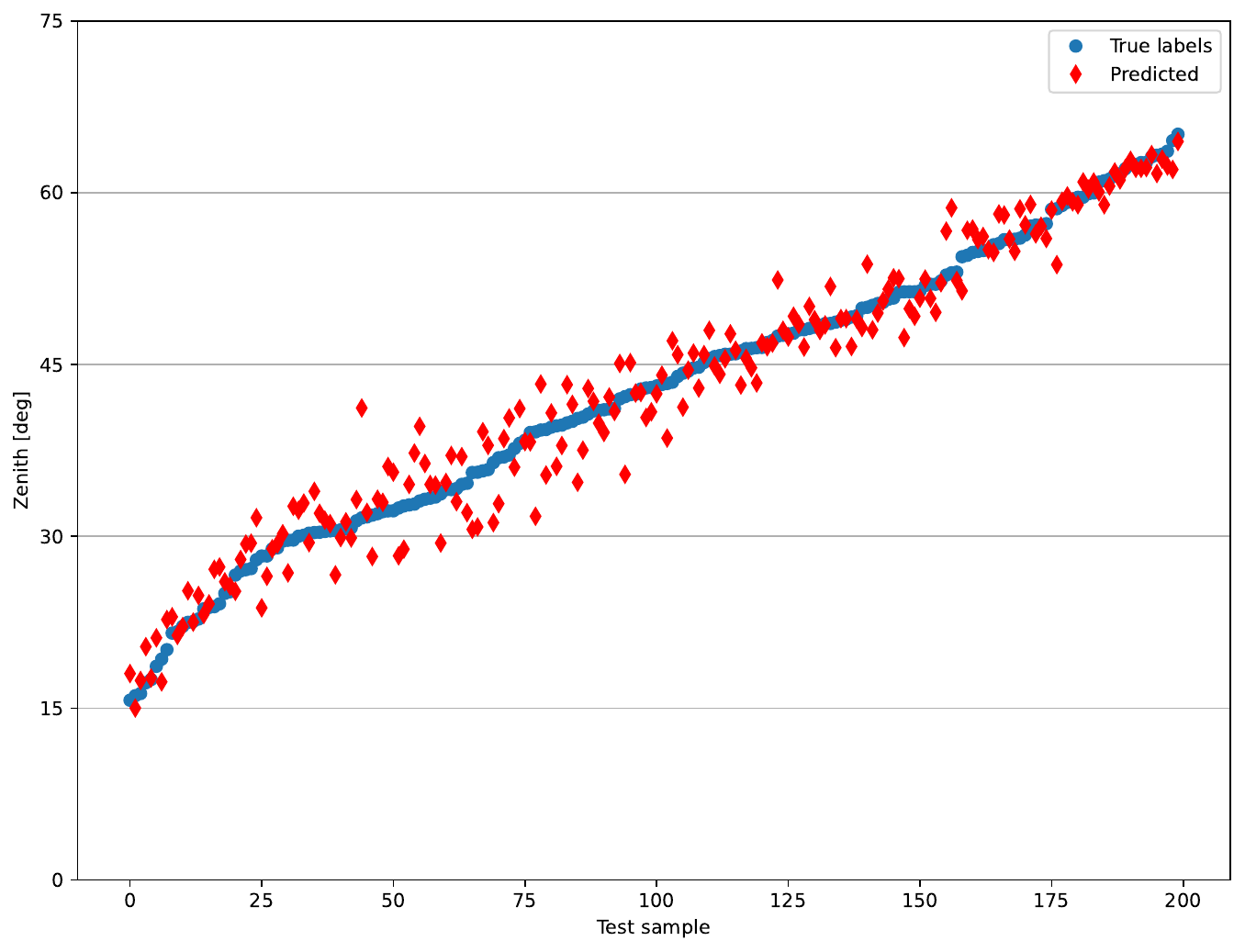}
	\caption{An example of reconstruction of zenith angles for a test
	sample for \spb.}
	\label{spb-zenith}
\end{figure}

Unexpectedly, zenith angles are reconstructed decently well. Probably
more surprising is the fact that mean errors become less if zenith
angles are reconstructed by themselves, as the only parameter of
regression.
Figure~\ref{spb-zenith} presents an example of reconstruction of zenith
angles performed separately from reconstruction of azimuth angles for
the same test sample. The mean error equals~$0.1^\circ$ with the
standard deviation~2.4.
The coefficient of determination equals~0.961.
The mean absolute error is~$1.85^\circ$.
The same error for zenith angles reconstructed simultaneously with
azimuth angles, was equal to~$3.17^\circ$.

It is interesting to notice that, according to our preliminary tests,
the above results can be improved if the input data is arranged in
stacks of ``screenshots'' of the focal surface instead of just integral
tracks.  One can say that in this case the CNN receives multi-channel
images instead of ``black-and-white'' ones.  This representation of data
puts higher demands on computer resources (mainly the size of memory)
but allows one to demonstrate kinematics of signals.  In particular, the
CNN trained to reconstruct ADs on data with 6 channels reduced errors of
angular reconstruction for the events shown in the left and right panels
of Fig.~\ref{spb-angsep-out} down to~$1.9^\circ$ and~$10.9^\circ$
respectively. However, some outliers remained for this data
representation, too.  We plan to study this in more detail in the
future.

Finally, let us remark that one can also reconstruct energies and
arrival directions simultaneously. This does not improve the overall
quality of the result but can reduce errors for events that were
outliers for reconstruction of energy. For example, the simultaneous
reconstruction of energy and ADs for the events shown in
Fig.~\ref{spb-energy-out} resulted in estimated energies of 44~EeV and
40.1~EeV respectively thus reducing errors to~24\% and~18\%
respectively, which are still large but better than with the initial
reconstruction.  Anyway, it is clear that tracks like these should be
analyzed especially carefully.

\subsection{EUSO-TA}

Reconstruction of energy and arrival directions of UHECRs registered
with a single ground-based fluorescence telescope is not a trivial task.

\begin{figure}[!ht]
	\includegraphics[height=.145\textheight]{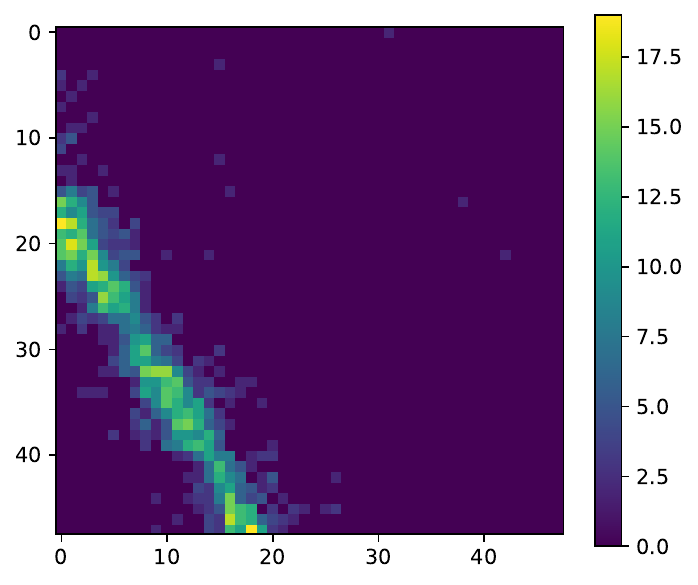}%
	\includegraphics[height=.145\textheight]{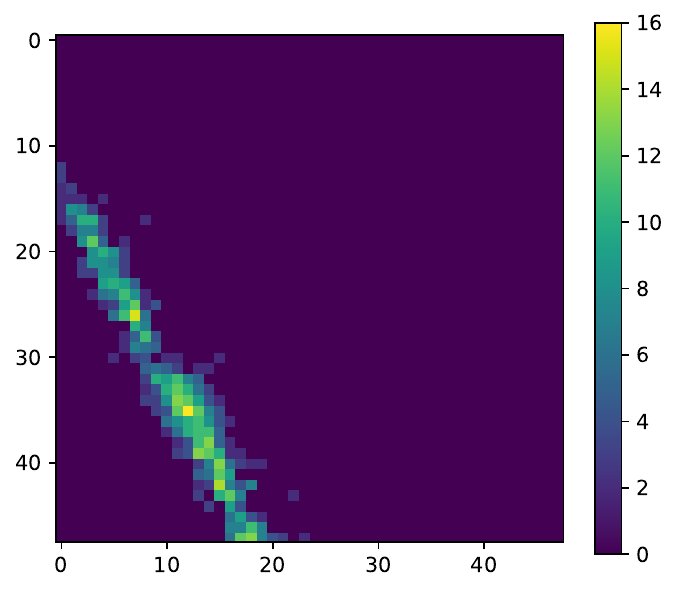}
	
	\caption{An example of two integral tracks simulated for \ta.
	They were produced by EASs originated from a 30~EeV (left) and
	a 98~EeV (right) protons. See the text for details.}

	\label{ta-tracks}
\end{figure}

One of the problems with reconstruction of energy of primary UHECRs
registered by the fluorescence radiation of their EASs with a
ground-based telescope, especially a single one,
is that the amplitude (luminosity) of the signal
strongly depends on the distance between the telescope and the axis of a
shower.  A shower originated from a lower energy particle but located
close to the detector can provide a brighter signal than a more distant
shower produced by a higher energy primary.
This problem is illustrated in Fig.~\ref{ta-tracks}.

The track shown in the left panel of Fig.~\ref{ta-tracks} was produced
by an EAS from a 30~EeV proton arriving at $(\theta, \phi)=(39^\circ, 134^\circ)$
with the shower core located in 10.3~km from the telescope.
(In the local coordinate system, \ta{} is pointed towards the azimuth angle
$\phi=0^\circ$.) 
The track shown in the right panel originated from a shower born by a
98~EeV proton that arrived from $(\theta, \phi)=(44^\circ, 37^\circ)$
and hit the ground in 21.6~km from the detector.
Signals of both events were non-zero during only three time steps.
Notice that the peak luminosity of the lower energy event is slightly
higher than that of the much more energetic one.

Another problem is specific to a small telescope with a narrow FoV, like
\ta. The point is that it is able to observe only a small part of a
shower, which results in incomplete information about the shower
luminosity and, thus, energy; see an in-depth discussion of this problem
in~\cite{eusota-2024}.
Thus it didn't come as a surprise when
our early attempts to reconstruct energy based on pure integral tracks
failed.  However, arranging input data in heaps of ``screenshots'' of
the focal surface greatly improved the situation. In what follows, we
present results obtained for data in which every event was presented by
12 ``screenshots'' of the FS made in consecutive moments of time.%
\footnote{Data with the time resolution of 1~$\mu$s (or another one)
might need another number of time steps to be used.}

\begin{figure}[!ht]
	\includegraphics[width=.48\textwidth]{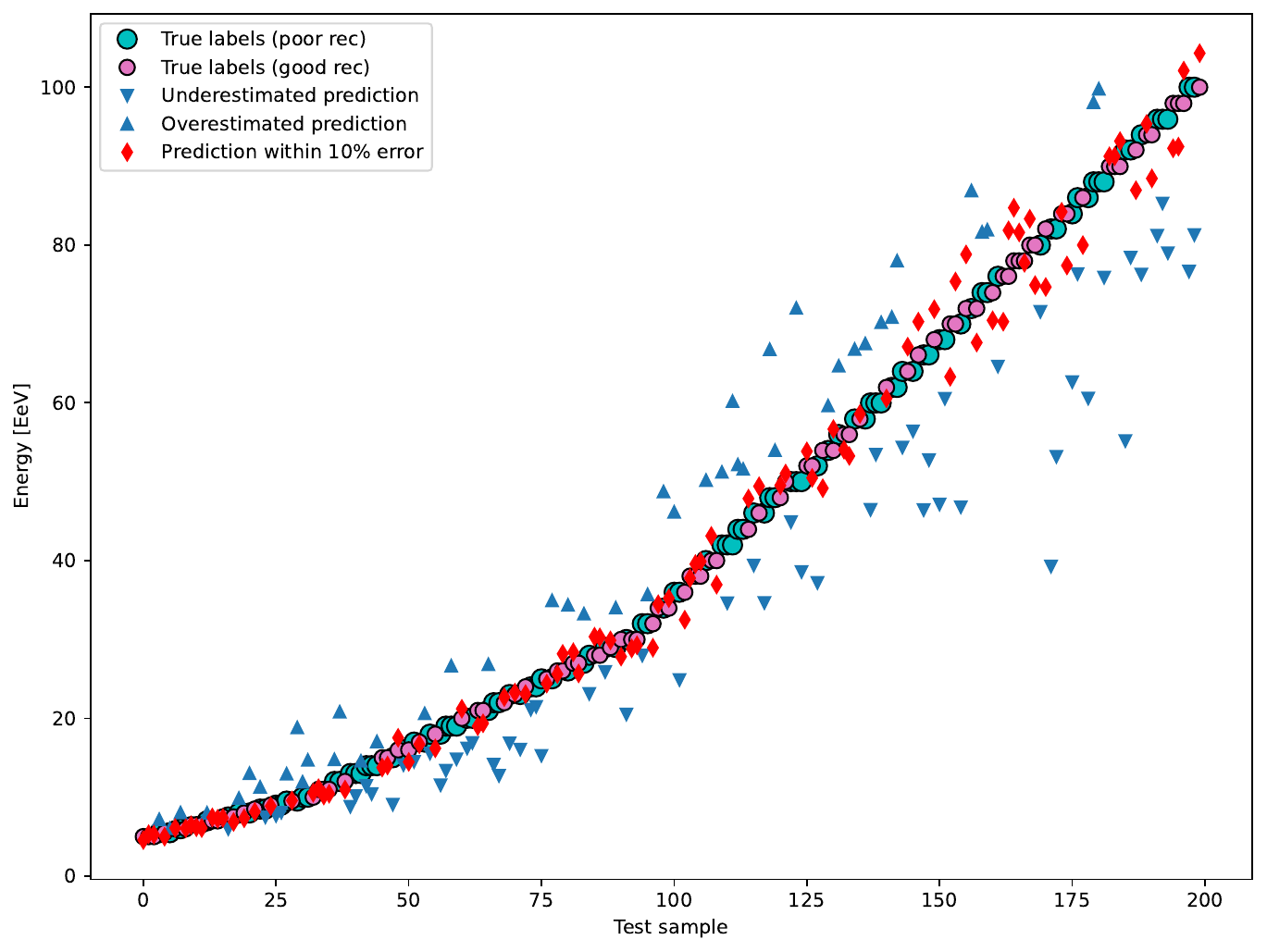}
	\caption{True (circles) and predicted values of energy for a test
	sample for \ta. Diamonds indicate predictions that deviate from true
	labels by less than~10\%.}
	\label{ta-energy}
\end{figure}

\begin{figure}[!ht]
	\includegraphics[width=.48\textwidth]{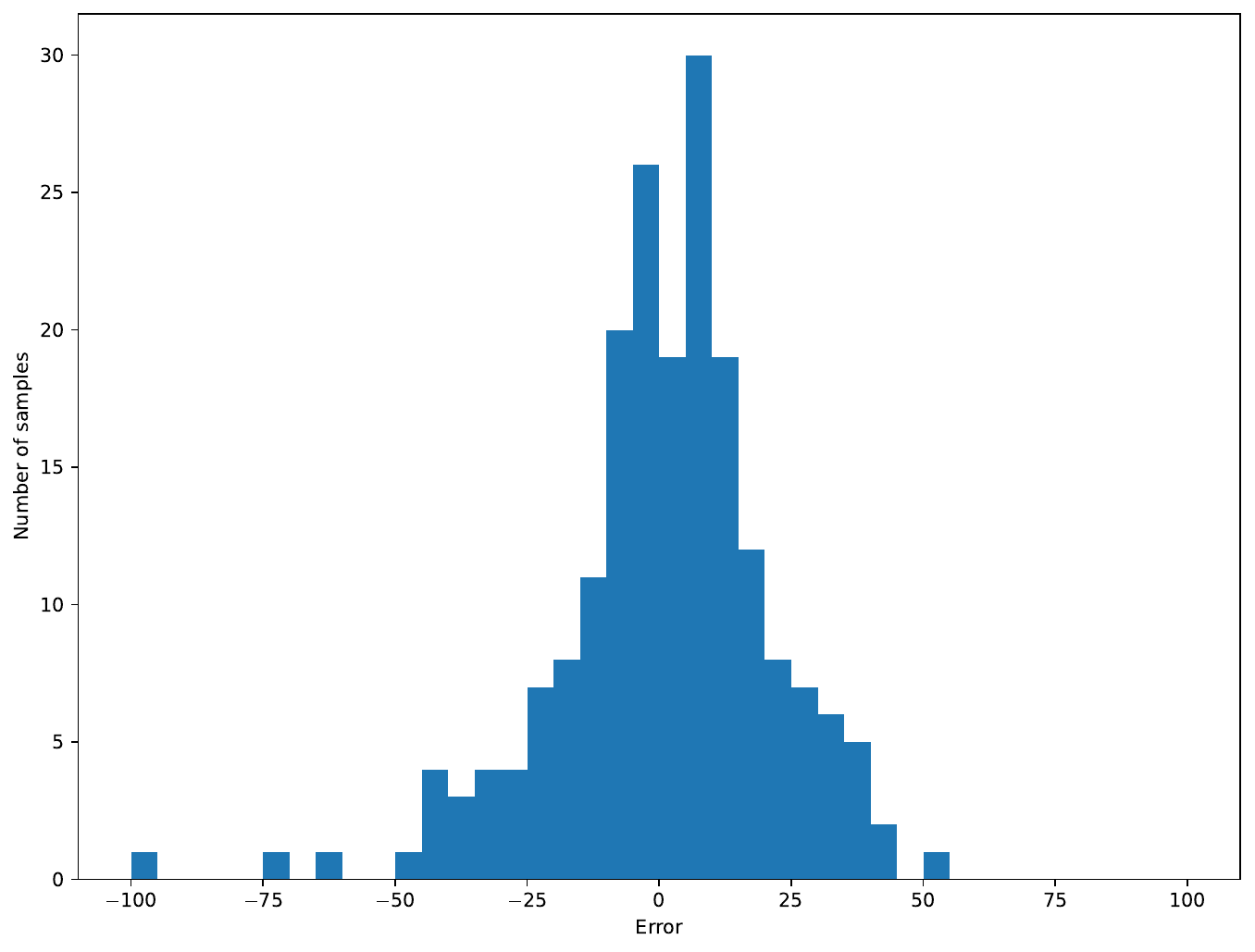}
	\caption{Distribution of errors of energy reconstruction for
	the test sample, expressed in percent.
	The mean equals~0.8, the standard deviation is equal to 20.7.}
	\label{ta-energy-hist}
\end{figure}

For \ta, we have extended the training and testing data sets down to
5~EeV in comparison with the \spb{} data discussed above. This
additional data set covering the range 5--10~EeV was simulated with
the step equal to 0.5~EeV. The training data set included 52 thousand events.
Figures~\ref{ta-energy} and~\ref{ta-energy-hist} present results of a
typical test on reconstruction of energy for \ta. In this case, the MAPE
equals 15.2\%.
Maximum error equals 98\% while the mean error is~0.8\% with the
standard deviation equal to 20.7\%.

\begin{figure}[!ht]
	\includegraphics[width=.24\textwidth]{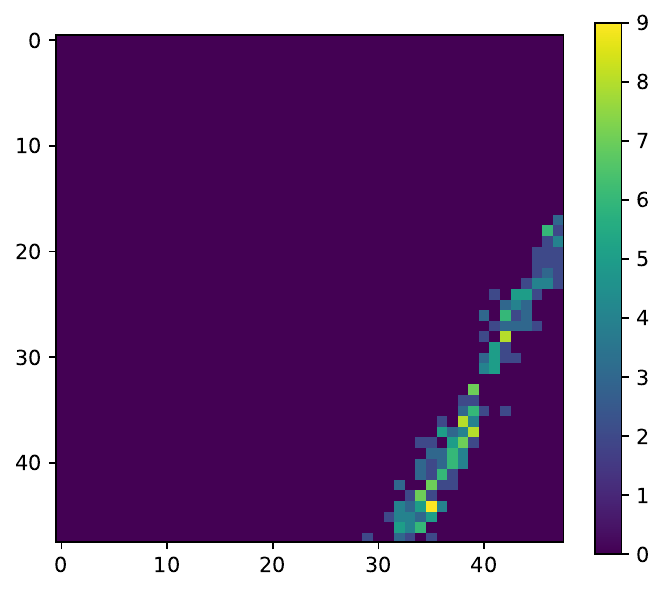}%
	\includegraphics[width=.24\textwidth]{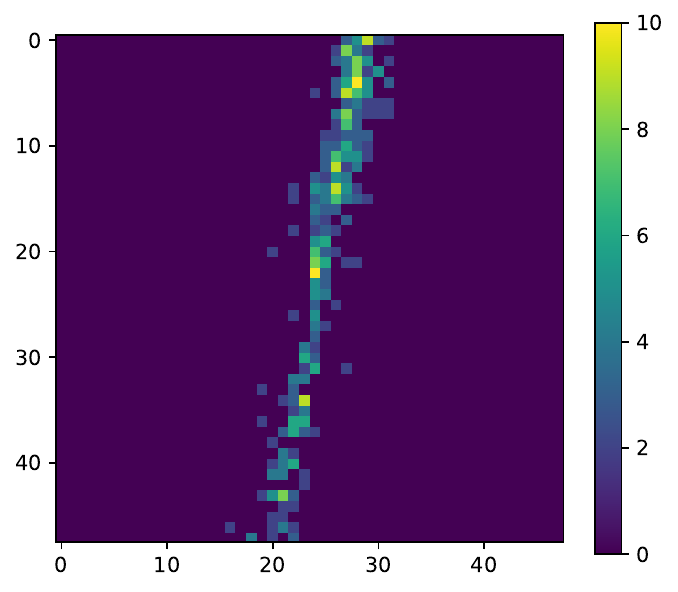}
	\caption{Integral tracks of two events simulated for \ta.
	See the text for details.}
	\label{ta-energy-out}
\end{figure}

The largest errors as expressed in percent take place for events with
lower energies, and their origin is not always clear.  Let us consider
as an example the event with the largest error in energy reconstruction.
Its integral track is shown in the left panel of
Fig.~\ref{ta-energy-out}. The event had true energy equal to 9.5~EeV
with the reconstructed value of 18.8~EeV resulting in a 98\% error.
The shower core was located in 12.1~km from the telescope. The arrival
direction in the local coordinate system was $(\theta,\phi)=(26^\circ,
274^\circ)$, i.e., it arrived in the direction almost orthogonal to the
axis of the field of view of \ta.

The integral track of another event is shown in the right panel of the
same figure. It has the same true energy, and the reconstructed energy
was estimated as 9.6~EeV resulting in mere 1\% error. The core of the
shower hit the ground in 14.1~km from the telescope. Both events have
nearly the same photon counts at their maxima and similar peak
luminosity. An important difference with the first event is that the
latter one arrived at the zenith angle equal to $51^\circ$ almost
parallel to the axis of the FoV but from behind the telescope
($\phi=193^\circ$). As a result, the signal was registered during~7 time
steps instead of two for the first event.

We remark that energy of both events shown in Fig.~\ref{ta-tracks}
was reconstructed with errors $\le6\%$.

We tried to reconstruct energy simultaneously with arrival
directions and with the distance from the telescope to the shower core
but this didn't improve the performance of the models.
Other preliminary tests demonstrated that errors can be reduced if the
whole energy range 5--100~EeV is split into several smaller ranges,
and the model is trained for each of them separately.

\begin{figure}[!ht]
	\includegraphics[width=.48\textwidth]{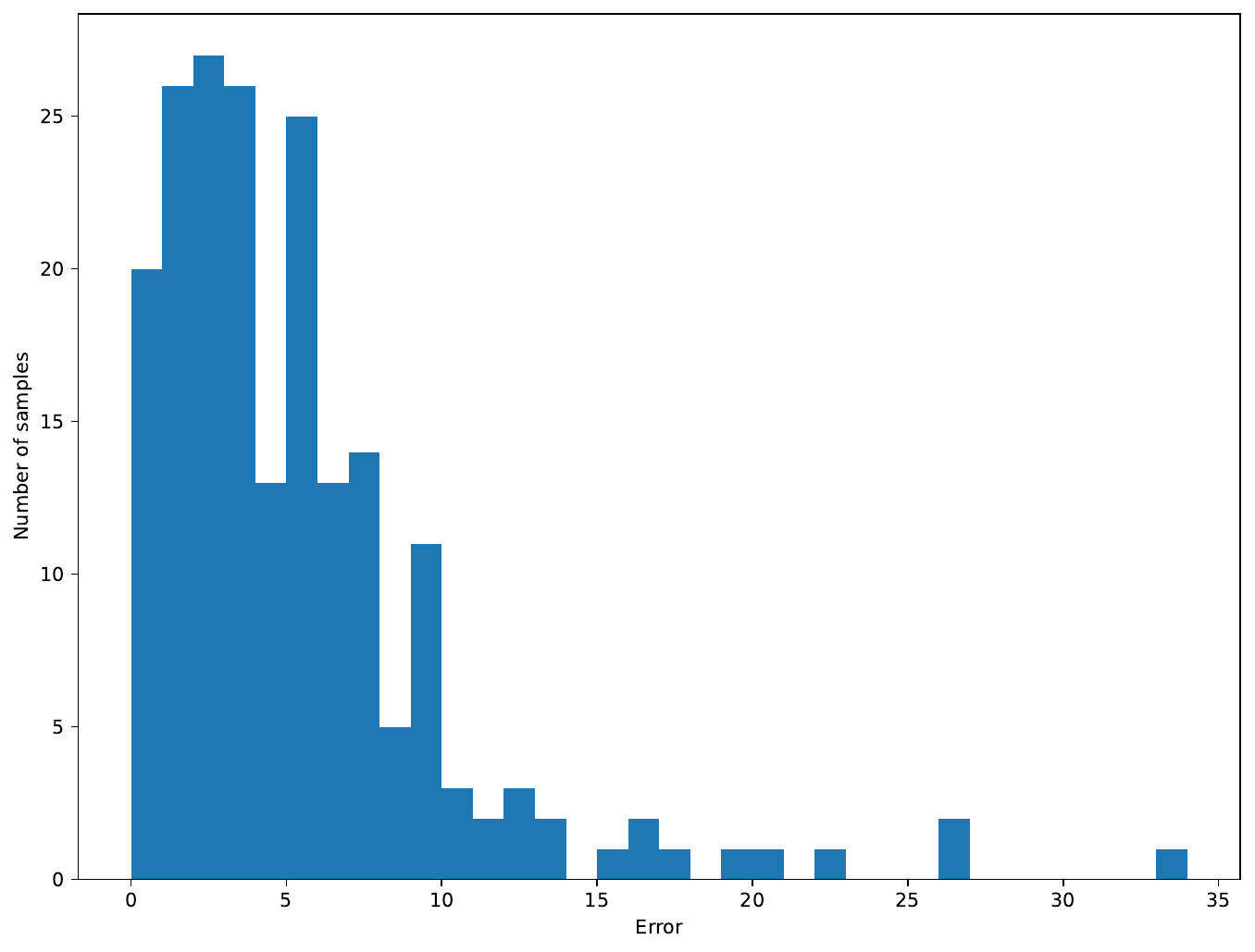}
	\caption{Distribution of angular separation between true and
	predicted arrival directions for a test sample for \ta.
	Values are expressed in degrees.}
	\label{ta-angsep}
\end{figure}

Now let us consider reconstruction of arrival directions of UHECRs
as they can be seen by \ta{} according to simulations.
Figure~\ref{ta-angsep} presents a histogram of errors in angular
separation between true and predicted arrival directions for the same test
sample consisting of 200 events with energies in the range 5--100~EeV.
The median value of the error equals~$4.1^\circ$, $R^2=0.913$.

\begin{figure}[!ht]
	\includegraphics[width=.48\textwidth]{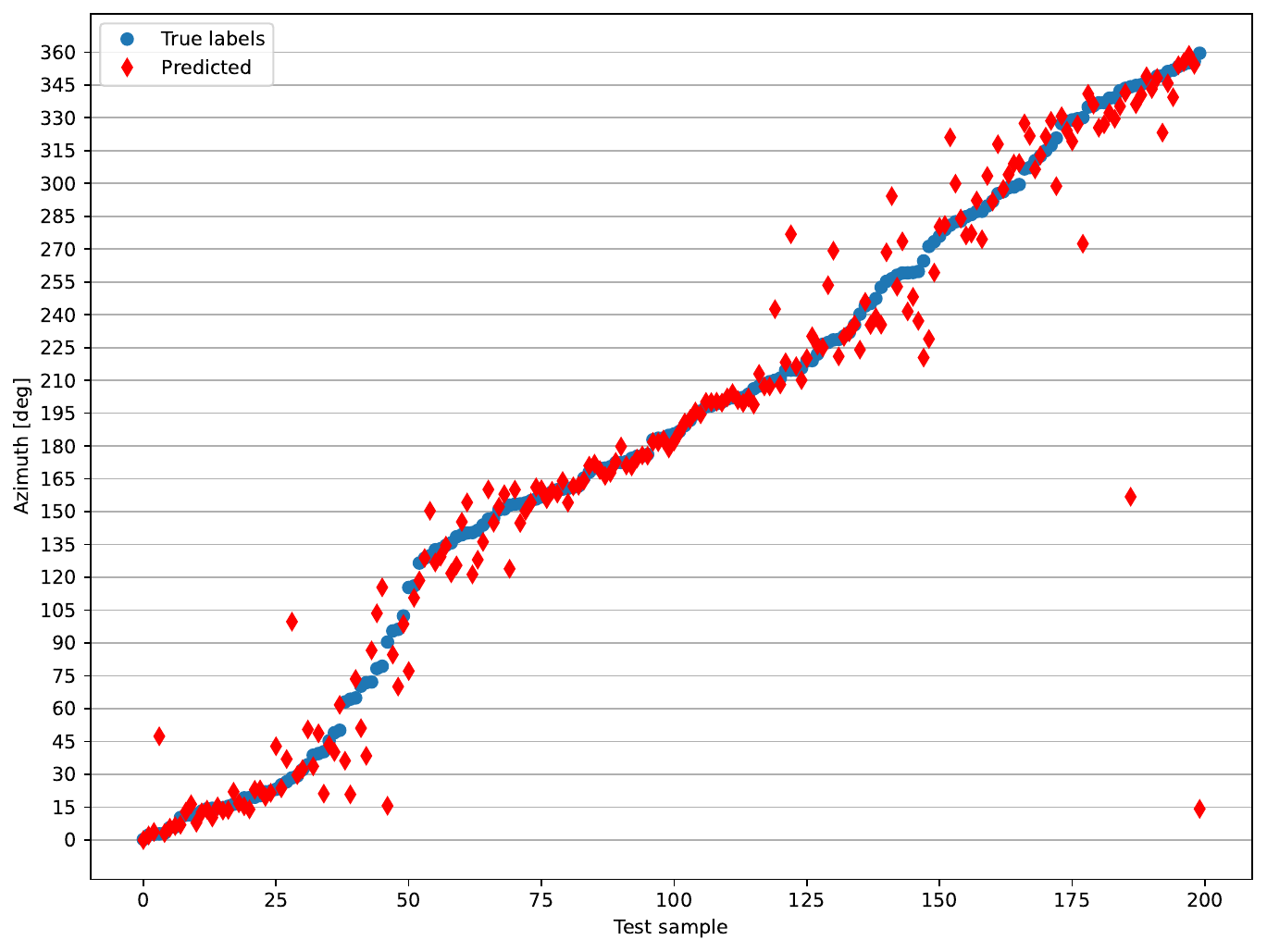}
	\caption{True (circles) and predicted (diamonds) values
	of azimuth angles for a test sample for \ta.}
	\label{ta-azimuth}
\end{figure}

\begin{figure}[!ht]
	\includegraphics[width=.48\textwidth]{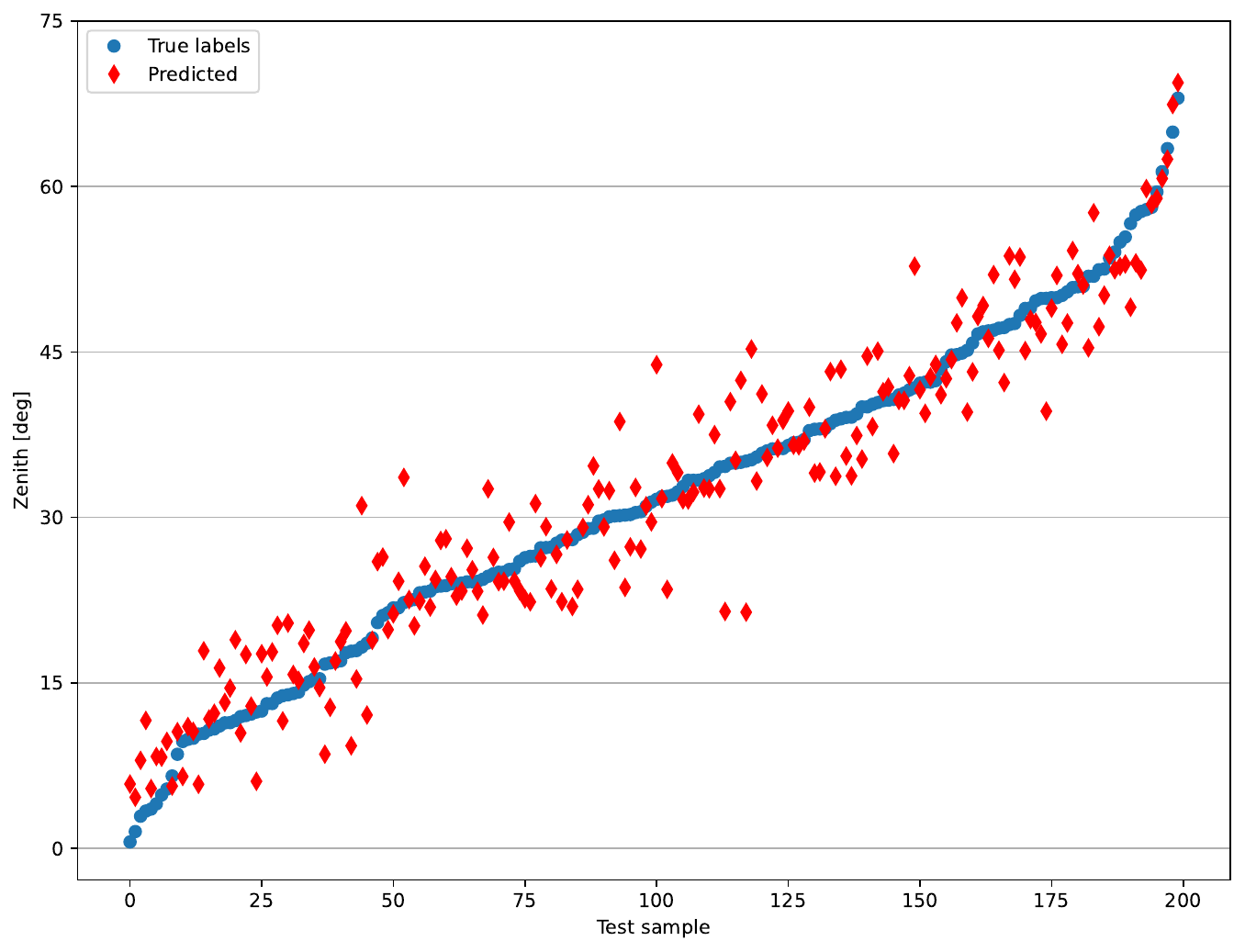}
	\caption{True (circles) and predicted (diamonds) values
	of zenith angles for a test sample for \ta.}
	\label{ta-zenith}
\end{figure}

Figures~\ref{ta-azimuth} and~\ref{ta-zenith} demonstrate how azimuth and
zenith angles were reconstructed in the above test. Notice that azimuth
angles around $0^\circ\pm20^\circ$ and $180^\circ\pm20^\circ$ are
mostly reconstructed pretty accurately. This direction corresponds
to the axis of \ta{} directed to $\phi=0^\circ$ in the local coordinate
system.

\begin{figure}[!ht]
	\includegraphics[width=.24\textwidth]{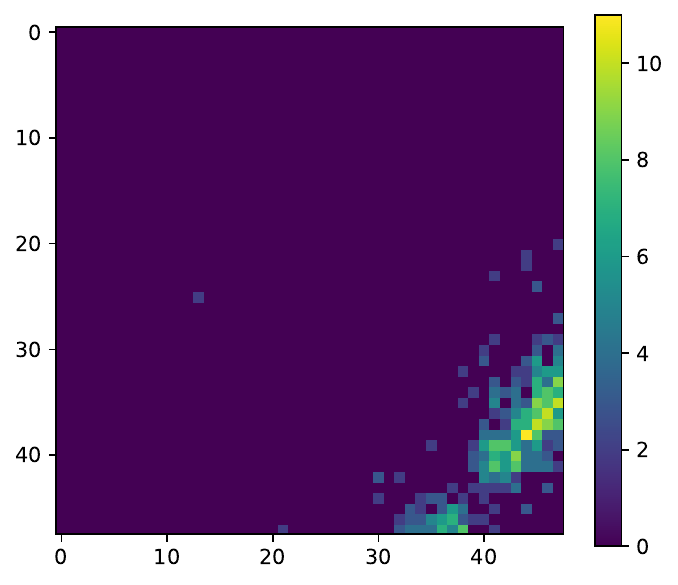}%
	\includegraphics[width=.24\textwidth]{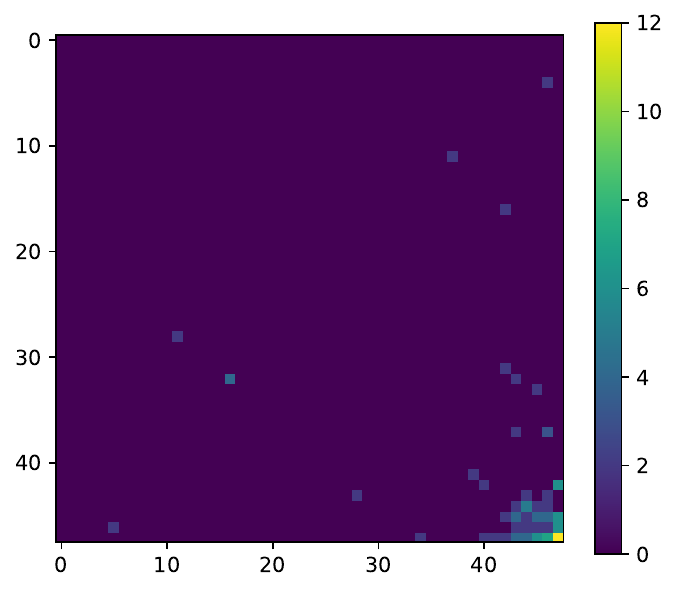}
	\caption{Integral tracks of events with the largest errors
	in arrival direction reconstruction. See the text for details.}
	\label{ta-angsep-out}
\end{figure}

Let us look at the two events with the largest errors in reconstruction
of their arrival directions. Their integral
tracks are presented in Fig.~\ref{ta-angsep-out}. The event shown in the
left panel has true $(\theta, \phi) = (38.9^\circ, 280.9^\circ)$ in the
local system of coordinates but $(\theta, \phi) = (43.4^\circ,
321.1^\circ)$ were predicted resulting in the angular separation equal
to $26.5^\circ$. The error might be due to the fact that the event had
only two time steps with nonzero photon counts. The track of the second
event touched only the corner of the field of view, resulting in an
error equal to $33.6^\circ$.

Our tests have revealed that models trained with slightly different
initial parameters sometimes result in noticeable difference in
predictions for ``difficult'' events.  When found, such events
can be treated with greater care.

\section{Track recognition}

It was assumed above that EAS tracks are recognized somehow, except dim
pixels with photon counts corresponding to the average rate of the
background illumination. Now let us consider how neural networks can be
employed to solve this task. One of the possible approaches is called
``semantic segmentation,'' which means that every pixel of an image
should be assigned to a certain class. We have just two classes of
pixels: those that form a track, and all the rest. Thus our task is to
assign the corresponding labels to all pixels as accurately as possible.

One of the established approaches is using a convolutional
encoder-decoder~\cite{segnet}. We implemented such a neural network
using eleven convolutional layers. Maxpooling layers were used after the
second and fourth CLs; upsampling was implemented after the 6th and 8th
CLs. The first nine CLs used 32 filters, the tenth one used 16 filters,
and one filter was used in the last layer. All convolutional layers
employed kernels of the size $3\times3$. Categorical cross-entropy was
used as the model loss function. Area under the precision-recall curve
(PR AUC) and binary cross-entropy were used as performance metrics
during model training.

Different functions can serve as a metric for evaluating the accuracy of
trained models. We have tried three such functions:
\begin{itemize}
	\item area under the precision-recall curve, where
	\[
		\text{Precision} = \frac{\tp}{\tp + \fp},\,
		\text{Recall} = \text{TPR} = \frac{\tp}{\tp + \fn};
	\]

	\item mean intersection-over-union (IoU), where
	\[
		\text{IoU} = \frac{\tp}{\tp+\fp+\fn};
	\]

	\item balanced accuracy
	$(\text{TPR} + \text{TNR})/2,$

\end{itemize}
where \tp, \fp, and \fn{} denote the number of true positives, false
positives, and false negatives respectively; TPR and TNR denote the true
positive and true negative rates. Recall that
the mean IoU function is a common evaluation metric for semantic image
segmentation, while balanced accuracy is useful, for example, in
classification tasks in which the number of positive samples is much
less than the number of negative ones, as is the case in our task.
In the perfect case, all three metrics are equal to~1.

Before trying to recognize a track, one has to decide how to label
pixels for training the model. In our approach, every data sample (an
integral track for \spb{} or a 12-layer data chunk for \ta), was scaled
to $[0, 1]$ before processing.  Then we considered how we should label
pixels on data samples without background illumination to reach the best
recognition of hit pixels in noisy data. We tested several thresholds
$\alpha=0.1,\dots,0.5$ assigning all pixels above the threshold a
value~1 and 0 to all the rest, training the model and then evaluating
the performance of track recognition.

\begin{figure}[!ht]
	\includegraphics[width=.48\textwidth]{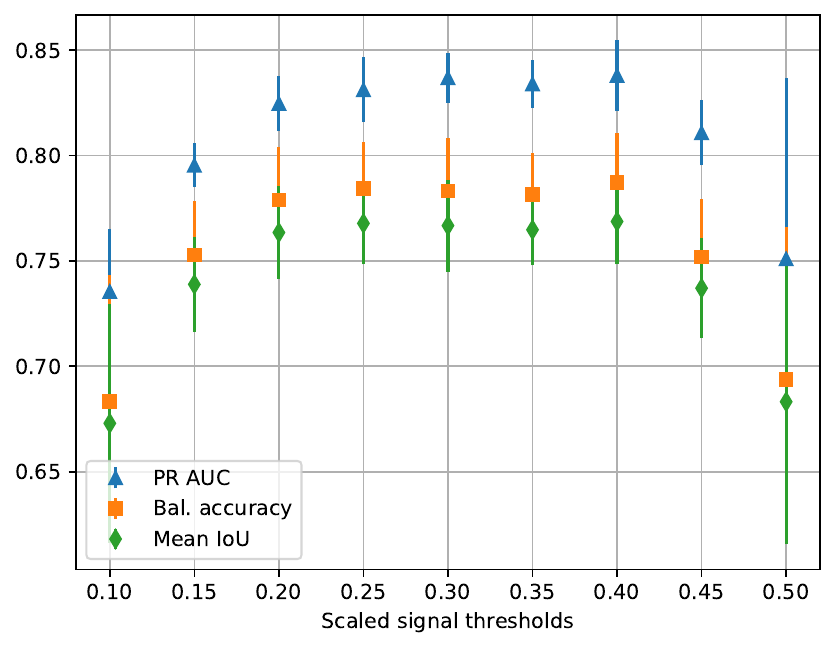}
	
	\caption{Performance metrics for track recognition with different
	thresholds used for labelling hit pixels. See the text for details.}

	\label{threshold}
\end{figure}

We considered samples simulated for the whole range of energies as well
as samples with a fixed energy of the primary particle.
Figure~\ref{threshold} presents results obtained for tracks generated by
50~EeV protons. The whole sample consisted of 14.6 thousand events.
Training of the model was repeated 10 times for each cut with different
initial weights.  For each run, the initial sample was split in a random
fashion into 1000 test events and 13.6 thousand events used for
training.  One can see that all three metrics demonstrate a local
maximum in the range of thresholds $\alpha\approx0.25,\dots,0.35$.
Similar results were obtained in other tests. Thus, in what follows we
mark hit pixels as belonging to a track if the value of the scaled
signal is above~0.25.

\begin{figure}[!ht]
	\includegraphics[width=.48\textwidth]{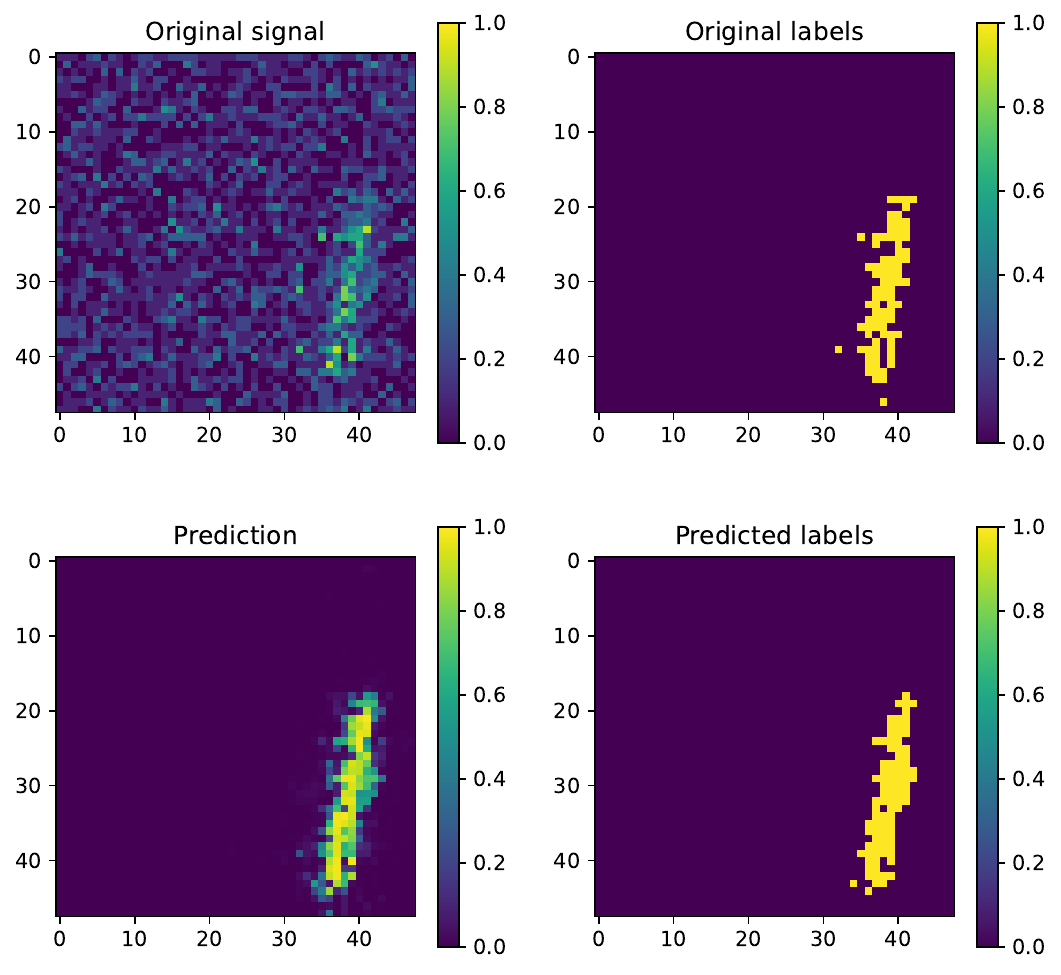}

	\caption{An example of track recognition for \ta.  The top row: the
	original signal (left) and the labels assigned using the signal
	without background illumination (right).  The bottom row:
	probabilities predicted by the neural network for the pixels to
	belong to the track (left) and the assigned labels (right).  See the
	text for details.}
	
	\label{ta-recognition}
\end{figure}

Let us illustrate how the implemented procedure works using a
``snapshot'' of the focal surface of \ta. The top left panel in
Fig.~\ref{ta-recognition} shows a signal recorded in one moment of time.
The top right panel shows how the track was marked using the pure
signal.  The bottom row presents the result of applying the model: the
left panel shows predicted probabilities of pixels to belong to the
track, and the right one shows the recognized track with the probability
threshold equal to~0.5.

The model used the same \ta{} data set as discussed above, generated for
proton primaries with energies in the range 5--100~EeV. The training
sample contained 32 thousand events, each including 12 ``snapshots'' of
the focal surface, thus resulting in 384 thousand images. Each pixel in
these images was labeled following the procedure described above.
The testing sample included 500 events with the same structure, i.e., it
consisted of 6000 images (some of which did not contain tracks).
The resulting PR AUC, mean IoU, and balanced accuracy metrics were equal
to 0.941, 0.882, and 0.934 respectively.

Interestingly, one can speed up training the model and slightly improve
performance if images that do not contain hit pixels or contain just a
few of them are filtered out from the training and testing samples. For
example, a model trained and tested on the same data set but with images
containing at least 24 hit pixels, demonstrated PR AUC, mean IoU, and
balanced accuracy 0.958, 0.898, and 0.949 respectively. Filtering out
images without or with just a few hit pixels is rather easy using the
total light curve of an event.

\section{Discussion}

We have presented two proof-of-concept neural networks that can be used to
recognize tracks of extensive air showers registered by fluorescence
telescopes and to reconstruct energy and arrival directions of primary
UHECRs.  It is clear that final results of reconstruction will be worse
than those presented above because we have considered a simplified
model.  In particular, our preliminary tests performed for the \spb{}
data demonstrated that mean absolute percentage error of energy
reconstruction increases by approximately 4\% if only hit pixels
recognized with mean IoU $\ge0.89$ are taken into consideration. 
Still, the accuracy of energy reconstruction remains comparable with
estimations for a much more sophisticated JEM-EUSO telescope obtained by
conventional methods~\cite{jemeuso-energy}.
In general, it is clear from the above results that special care
should be taken for events that have just a few hit pixels, those
with tracks touching only the very edge of the focal surface or
located along the gaps between PDMs (in the case of \spb).
However, this is equally true for conventional methods.

We see a number of ways in which the results presented above can be
improved (besides tuning hyper-parameters of the neural networks):
\begin{itemize}

	\item One can reconstruct energy using smaller intervals and larger
		training data sets. Our preliminary tests demonstrated that this can
		decrease the MAPE by about~2\%.
		Splitting training sets into subsets with different ranges of
		zenith angles might also help.
		Specially crafted data sets with more representatives of
		quasi-vertical and quasi-horizontal air showers can be used to
		optimize reconstruction of arrival directions.

	\item We put very loose cuts on the signals (tracks) used for
		training and testing the neural networks but stricter cuts can
		considerably improve their performance. For example, we did not
		demand that the shower maximum is in the FoV of a telescope for
		incomplete tracks, which is often the case for \ta{} but also
		takes place for \spb. Our tests have demonstrated that if we
		demand the brightest pixel of a track to be in at least 4 pixels
		from the edge of the FoV of \spb, this decreases the MAPE for
		energy reconstruction by $\approx2\%$.

	\item One can simulate a larger field of view than that of an
		actual telescope and use it to teach a neural network to restore
		missing parts of a track. This technique is called image
		inpainting~\cite{inpainting}. It can be especially useful for
		fluorescence telescopes with a small FoV but large instruments can
		also benefit from it by reconstructing incomplete tracks.

\end{itemize}

We have not implemented a complete pipeline that includes both suggested
neural networks and have not tested it on real experimental data of
\ta{} yet.  This is a work in progress that will be reported elsewhere.

\begin{acknowledgments}
	We thank heartfully Francesca~Bisconti, George~Filippatos and
	Zbigniew~Plebaniak for their invaluable help with simulating data for
	both EUSO-SPB2 and EUSO-TA, and Mario~Bertaina for important comments
	on the manuscript.

	All models were implemented in Python using TensorFlow~\cite{tf}
	and scikit-learn~\cite{sklearn} libraries.
\end{acknowledgments}

\section*{FUNDING}
The development of neural networks for EUSO-SPB2 was partially supported by the
Russian Science Foundation grant 22-22-0367.
Simulations and the development of neural networks for EUSO-TA is
supported by the RSF grant 22-62-00010. 

\section*{CONFLICT OF INTEREST}
The authors declare that they have no conflicts of interest.

\subsection*{Publisher's Note}

Pleiades Publishing remains neutral with regard to jurisdictional claims
in published maps and institutional affiliations.

\end{document}